\documentclass[twocolumn,prb,amsfonts,amsmath,amssymb,floatfix]{revtex4} 
\usepackage{color}
\usepackage{hhline}
\usepackage{mathrsfs}
\usepackage{graphicx}
\usepackage{dcolumn}
\usepackage{bm}
\usepackage{multirow}
\usepackage{booktabs}
\usepackage{afterpage}
\usepackage{amsmath}
\usepackage{tabularx}

\begin{document}

\title{Comparative first-principles study of antiperovskite oxides and nitrides as thermoelectric material: multiple Dirac cones, low-dimensional band dispersion, and high valley degeneracy}

\author{Masayuki Ochi}
\author{Kazuhiko Kuroki}
\affiliation{Department of Physics, Osaka University, Machikaneyama-cho, Toyonaka, Osaka 560-0043, Japan}

\date{\today}
\begin{abstract}
We perform a comparative study on thermoelectric performance of antiperovskite oxides $Ae_3Tt$O and nitrides $Ae_3Pn$N ($Ae=$ Ca, Sr, Ba; $Tt=$ Ge, Sn, Pb; $Pn=$ As, Sb, Bi) by means of first-principles calculation.
As for the oxides with the cubic structure, Ca$_3$GeO with a sizable band gap exhibits high thermoelectric performance at high temperatures, while Ba$_3$PbO with Dirac cones without the gap is favorable at low temperatures. The latter high performance owes to high valley degeneracy including the multiple Dirac cones and the valleys near the $\Gamma$ and R points.
For the nitrides with the cubic structure, insulator with strong quasi-one-dimensionality exhibits high thermoelectric performance. We also find that the orthorhombic structural distortion sometimes sizably enhances thermoelectric performance, especially for Ba$_3$GeO and Sr$_3$AsN where the high valley degeneracy is realized in the $Pnma$ phase. Our calculation reveals that antiperovskites offer a fertile playground of various kinds of characteristic electronic structure, which enhance the thermoelectric performance, and provides promising candidates of high-performance thermoelectric materials.
\end{abstract}

\maketitle

\section{Introduction}
Searching high-performance thermoelectric materials is a central issue in the study of themoelectrics.
There are many promising compounds such as Bi$_2$Te$_3$~\cite{Bi2Te3_1,Bi2Te3_2,Bi2Te3_3}, lead chalcogenides~\cite{PbCh_1,PbCh_2,PbCh_3,PbCh_4}, skutterudites~\cite{sk_1,sk_2,sk_3,sk_4,sk_5}, clathrates~\cite{clathrates}, and Na$_x$CoO$_2$~\cite{NaxCoO2}.
These high-performance materials have some characteristics in their crystal and/or electronic structures.
For example, rattling motion of atoms is a key for low thermal conductivity in skutterudites and clathrates~\cite{clathrates,rattling1,rattling2,rattling3,rattling4,rattling5,rattling6,rattling7,rattling8,tadano1,tadano2}.
Band convergence studied in lead chalcogenides~\cite{band_conv_PbCh} is one of the most important and general concepts to enhance thermoelectric performance.
Low-dimensionality~\cite{Hicksone,Hicksone2,usuione,Fukuyama} is also an important concept for enhancing thermoelectric performance of many materials including layered materials, nanowires, and nanotubes~\cite{nano1,CNT1,CNT2}.
It was pointed out that high thermoelectric performance of Na$_x$CoO$_2$ originates from a pudding-mold-shaped band structure~\cite{pudding}, where a large group velocity and a high density of states (DOS) can coexist.
These studies show that investigating characteristic crystal and/or electronic structures often bring ones general and useful concepts for seeking high thermoelectric performance.

\begin{figure}
\begin{center}
\includegraphics[width=8.5 cm]{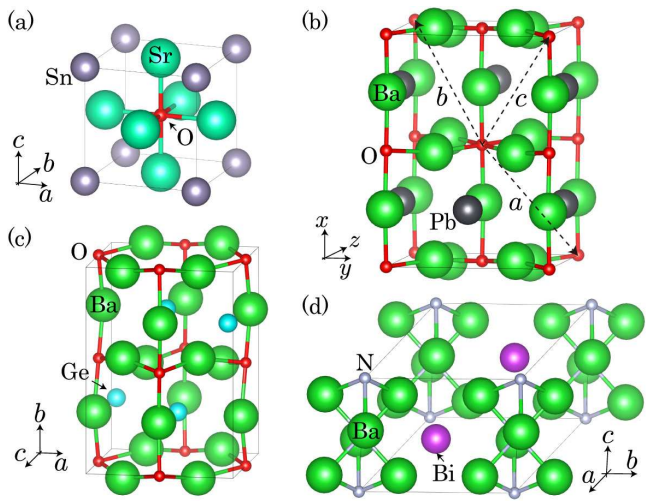}
\caption{Crystal structures of (a) Sr$_3$SnO ($Pm\bar{3}m$), (b) Ba$_3$PbO ($Imma$), (c) Ba$_3$GeO ($Pnma$), and (d) Ba$_3$BiO ($P$6$_3$/$mmc$). A doubled unit cell is shown in panel (b), while the lattice vectors $a$, $b$, and $c$ for the primitive unit cell are also shown with broken liens.
Depicted using the VESTA software~\cite{VESTA}.}
\label{fig:crys}
\end{center}
\end{figure}

Antiperovskites, in which the positions of constituent elements in a famous perovskite structure are interchanged as shown in Fig.~\ref{fig:crys}(a), have attracted much attention from several aspects, such as superconductivity~\cite{SC_MgCNi3,SC_Sr3SnO},
giant negative thermal expansion~\cite{nega_therm_exp1,nega_therm_exp2},
giant magnetoresistance~\cite{GMR}, magnetostriction~\cite{Mstriction}, and magnetocaloric effects~\cite{Mcaloric}.
Interestingly also, recent studies pointed out that some antiperovskite oxides and nitrides are candidates for three-dimensional massless Dirac electron systems~\cite{topo1,topo2,topo3} and topological crystalline insulators~\cite{topo4_TCI}.
They also belong to mixed-anion compounds~\cite{mixed_review}, which are characterized by multiple anion atoms. For example, Sr$_3$SnO has two kinds of anion atoms, O$^{2-}$ and Sn$^{4-}$, the latter of which is an unusual negative oxidation state of group-14 elements.
The unique crystal and electronic structures of antiperovskites have been investigated also as possible candidates for thermoelectric materials in experimental~\cite{Okamoto} and theoretical~\cite{Haddadi1,Haddadi2,Haddadi3,Bilal1,Bilal2,Hassan,Hassan2,Batool,Haque1,Haque2,Haque3,Haque4,Haque5} studies.
Experimental realization of the carrier control and a high Seebeck coefficient of around 100 $\mu$VK$^{-1}$ together with a metallic resistivity and a relatively low thermal conductivity of around 2 Wm$^{-1}$K$^{-1}$ at room temperature is promising~\cite{Okamoto}.
However, it is still unclear whether their characteristic electronic structure including the Dirac dispersion is favorable for thermoelectric performance and how to enhance their performance.
Because of their unique characteristics, it is expected that investigation on the thermoelectric properties of antiperovskites will provide novel and important knowledge that will expand the possibility for further findings of high-performance thermoelectric materials.

In this paper, we perform a comparative study on thermoelectric performance of antiperovskite oxides $Ae_3Tt$O and nitrides $Ae_3Pn$N ($Ae=$ Ca, Sr, Ba; $Tt=$ Ge, Sn, Pb; $Pn=$ As, Sb, Bi) by means of first-principles calculation.
We find that several kinds of characteristic electronic structure play an important role in enhancing their thermoelectric performance: multiple Dirac cones, quasi-one-dimensional band dispersion, and high valley degeneracy induced by the structural distortion toward the orthorhombic $Pnma$ phase.
Here, because the crystal structure of the antiperovskite oxides and nitrides exhibits some variations as presented in Tables~\ref{table:crys_o} and \ref{table:crys_n}, we investigate the effect of the structural change onto their thermoelectric performance.
Our study reveals a unique and fertile electronic structure of the antiperovskite oxides and nitrides, which is attracting also as thermoelectric materials,
and provides possible promising candidates for high-performance thermoelectric materials.

\begin{table}
\caption{\label{table:crys_o} Space group of each oxide $Ae_3Tt$O ($Ae=$ Ca, Sr, Ba; $Tt=$ Ge, Sn, Pb) in experiments.
(ht) and (lt) denote the high-temperature and low-temperature phases, respectively.
Information for Ba$_3$GeO and Sr$_3$GeO were taken from Refs.~[\onlinecite{oxides0}] and [\onlinecite{oxides1}], respectively, and others were taken from Ref.~[\onlinecite{oxides2}]. A transition temperature of Ca$_3$GeO, Ba$_3$SnO, and Ba$_3$PbO, was reported to be around 350 K, around 150 K, and around 150 K, respectively~\cite{oxides2}.}
\begin{center}
\scalebox{0.9}{
\begin{tabular}{c c c c}
\hline \hline
$Tt$$\backslash$$Ae$& Ca & Sr & Ba \\
\hline
Ge & $Pm\bar{3}m$ (ht), $Imma$ (lt) & $Pnma$ & $Pnma$ \\
Sn & $Pm\bar{3}m$ & $Pm\bar{3}m$ &$Pm\bar{3}m$ (ht), $Imma$ (lt)  \\
Pb & $Pm\bar{3}m$ & $Pm\bar{3}m$ & $Pm\bar{3}m$ (ht), $Imma$ (lt) \\
\hline \hline
\end{tabular}
}
\end{center}
\end{table}

\begin{table}
\caption{\label{table:crys_n} Space group of each nitride $Ae_3Pn$N ($Ae=$ Ca, Sr, Ba; $Pn=$ As, Sb, Bi) in experiments.
(ht) and (lt) denote the high-temperature and low-temperature phases, respectively.
Information for $Pn=$ Sr and Ba was taken from Ref.~[\onlinecite{nitrides3}] and that for $Pn=$ Ca was taken from Refs.~[\onlinecite{nitrides4,nitrides5}].
A transition temperature of Ca$_3$AsN was reported to be 1025 K~\cite{nitrides4}.}
\begin{center}
\begin{tabular}{c c c c}
\hline \hline
$Pn$$\backslash$$Ae$ & Ca & Sr & Ba \\
\hline
As & $Pm\bar{3}m$ (ht), $Pnma$ (lt) & - & - \\
Sb & $Pm\bar{3}m$ & $Pm\bar{3}m$ &$P$6$_3$/$mmc$ \\
Bi & $Pm\bar{3}m$ & $Pm\bar{3}m$ & $P$6$_3$/$mmc$ \\
\hline \hline
\end{tabular}
\end{center}
\end{table}

This paper is organized as follows.
Section~\ref{sec:method} presents calculation methods we employed in this study.
Our calculation results for the antiperovskite oxides and nitrides with the cubic ($Pm\bar{3}m$) structure are presented in Sec.~\ref{sec:o} and \ref{sec:n}, respectively.
In Sec.~\ref{sec:struct}, thermoelectric performance of the antiperovskite oxides and nitrides with the orthorhombic ($Imma$ and $Pnma$) and hexagonal ($P6_3/mmc$) structures are discussed.
This study is summarized in Sec.~\ref{sec:sum}.
 
\section{Calculation methods~\label{sec:method}}

First, we performed the structural optimization using the PBEsol exchange-correlation functional~\cite{PBEsol} and the projector augmented wave method~\cite{paw} with the inclusion of the spin-orbit coupling (SOC). For this purpose, we used {\it Vienna ab initio Simulation Package} (VASP)~\cite{vasp1,vasp2,vasp3,vasp4}.
For the $Pm\bar{3}m$, $Imma$, $Pnma$, and $P$6$_3$/$mmc$ space groups, 12$\times$12$\times$12, 10$\times$10$\times$10,  10$\times$6$\times$10, and 12$\times$12$\times$12 $\bm{k}$-meshes were used, respectively. Crystal structures of antiperovskites with these space groups are shown in Fig.~\ref{fig:crys}.
A plane-wave cutoff energy of 550 eV was used for all the cases.

After the structural optimization, we performed first-principles band-structure calculation using \textsc{WIEN2k} code~\cite{wien2k}. We employed the Tran-Blaha modified Becke-Johnson (TB-mBJ) potential~\cite{mBJ1,mBJ2} to obtain a reliable size of the band gap.
In self-consistent-field (SCF) calculations for the $Pm\bar{3}m$, $Imma$, $Pnma$, and $P$6$_3$/$mmc$ space groups, 
12$\times$12$\times$12, 10$\times$10$\times$10, 8$\times$6$\times$8, and 10$\times$10$\times$10 $\bm{k}$-meshes were used, respectively.
For calculating DOS, we took 54$\times$54$\times$54 and 38$\times$27$\times$38 $\bm{k}$-meshes for the $Pm\bar{3}m$ and $Pnma$ space groups, respectively.
We used a relatively high value of the $RK_{\mathrm{max}}$ parameter, 10, since Wannier functions in a high energy region were extracted as mentioned below. SOC was included unless noted.

From the calculated band structures, we extracted the Wannier functions of the $Ae$-$d$, $Tt$($Pn$)-$p$, and O(N)-$p$ orbitals using the \textsc{Wien2Wannier} and \textsc{Wannier90} codes~\cite{Wannier1,Wannier2,Wien2Wannier,Wannier90}.
We did not perform the maximal localization procedure for the Wannier functions to prevent orbital mixing among the different spin components.
For the $Pm\bar{3}m$, $Imma$, $Pnma$, and $P$6$_3$/$mmc$ space groups, 
we used $16\times 16\times 16$, 12$\times$12$\times$12, 12$\times$8$\times$12, and 10$\times$10$\times$10 $\bm{k}$-meshes, respectively, for constructing the Wannier functions.
Then, we constructed the tight-binding model with the obtained hopping parameters among the Wannier functions.
We analyzed the transport properties using this model with the Boltzmann transport theory.
The transport coefficients ${\bf K}_{\nu}$ are represented as follows:
\begin{align}
{\bf K}_\nu= \tau\sum_{n,{\bm{k}}} \bm{v}_{n,\bm{k}}\otimes\bm{v}_{n,\bm{k}}\left[-\frac{\partial f_0}{\partial \epsilon_{n,\bm{k}}}\right](\epsilon_{n,\bm{k}}-\mu(T))^\nu ,\label{eq:transp}
\end{align}
with the Fermi--Dirac distribution function $f_0$, chemical potential $\mu(T)$, energy $\epsilon_{n,\bm{k}}$ and group velocity $\bm{v}_{n,\bm{k}}$ of the one-electron orbital on the $n$-th band at the $\bm{k}$-point $\bm{k}$ and the relaxation time $\tau$, which was assumed to be constant in this study.
By using ${\bf K}_{\nu}$, the electrical conductivity ${\boldsymbol \sigma}$, Seebeck coefficient ${\bf S}$, and electrical thermal conductivity ${\boldsymbol \kappa}_{\rm el}$ are expressed as follows:
\begin{align}
{\boldsymbol \sigma}=e^2{\bf K}_0,\ \ \  {\bf S}=-\frac{1}{eT}{\bf K}_0^{-1}{\bf K}_1, \label{eq:sigmas}\\ 
{\boldsymbol \kappa}_{\rm el}=\frac{1}{T}\left[{\bf K}_2-{\bf K}_1{\bf K}_0^{-1}{\bf K}_1 \right] ,
\end{align}
where $e$ ($>0$) is the elementary charge. The power factor PF $=\sigma S^2$ and the dimensionless figure of merit $ZT=\sigma S^2 T \kappa^{-1}$ were also calculated using these quantities.
We assumed that the thermal conductivity ${\boldsymbol \kappa}$ can be represented as the sum of the electrical thermal conductivity ${\boldsymbol \kappa}_{\rm el}$ and the lattice electrical thermal conductivity ${\boldsymbol \kappa}_{\rm lat}$, namely, ${\boldsymbol \kappa} = {\boldsymbol \kappa}_{\rm el}+{\boldsymbol \kappa}_{\rm lat}$.
In our study, $\tau$ and $\kappa_{\rm lat}$ were assumed to be $10^{-14}$ second and 2 Wm$^{-1}$K$^{-1}$, respectively, which are typical values for thermoelectric materials.
In fact, the thermal conductivity of Ca$_3$SnO was reported to be around 2 Wm$^{-1}$K$^{-1}$ at room temperature~\cite{Okamoto}.
The relaxation time and the lattice thermal conductivity are generally different among materials, and so their theoretical evaluation based on phonon calculation is an important future issue.
In this study, we concentrated on how favorable the electronic band structure of candidate materials is.
We shall discuss this point in more detail through the comparison with the experimental results in Sec.~\ref{sec:TEO}.

To simulate the carrier doping, we adopted the rigid band approximation.
Because an ARPES experiment on the Dirac semimetallic Ca$_3$PbO~\cite{ARPES_Ca3PbO} reported that
the band structure for hole-doped Ca$_3$PbO, to say, Ca$_3$Pb$_{0.92}$Bi$_{0.08}$O, shows a very good agreement with
the calculated band structure for the mother compound Ca$_3$PbO,
we can expect that the rigid-band approximation is valid against a certain level of the carrier doping, even for the narrow- or zero-gap systems.
We only considered the hole carrier doping, which was realized in experiments for some antiperovskite oxides~\cite{SC_Sr3SnO,Okamoto}.
We employed a fine $\bm{k}$-mesh up to 900$\times$900$\times$900 for calculating the transport properties with sufficient convergence.

\section{Results and Discussions}

\subsection{Oxides $Ae_3Tt$O with the cubic structure~\label{sec:o}}

In this section, we investigated the electronic structure and the transport properties of the antiperovskite oxides $Ae_3Tt$O ($Ae=$ Ca, Sr, Ba; $Tt=$ Ge, Sn, Pb), assuming the cubic structure with a space group $Pm\bar{3}m$ for all the compounds.
Because some oxides have a distorted crystal structure as shown in Table~\ref{table:crys_o}, we shall see the effect of the structural (orthorhombic) distortion on thermoelectric performance in Sec.~\ref{sec:struct}.

\subsubsection{Band structures}

\begin{figure}
\begin{center}
\includegraphics[width=8.5 cm]{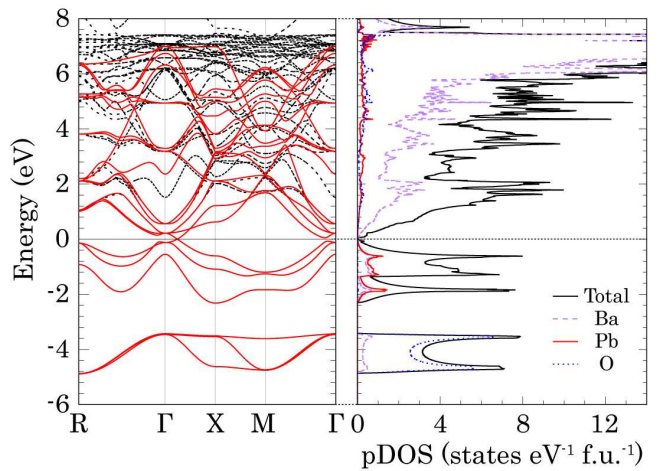}
\caption{Band structure and partial DOS (pDOS) of Ba$_3$PbO with the cubic structure (space group: $Pm\bar{3}m$). Black broken and red solid lines in the band structure represent the band structures obtained with the first-principles calculation and the tight-binding model for the Wannier functions, respectively.}
\label{fig:band_ba3pbo}
\end{center}
\end{figure}

Figure~\ref{fig:band_ba3pbo} presents a calculated band structure and partial DOS (pDOS) of Ba$_3$PbO with the cubic structure (space group: $Pm\bar{3}m$) as a typical member of the antiperovskite oxides.
Black broken and red solid lines in the band structure represent the band structures obtained with the first-principles calculation and the tight-binding model for the Wannier functions, respectively. As is clearly seen, the band structure calculated with the tight-binding model well reproduces the first-principles one near the Fermi energy, which validates our approach using the tight-binding model to evaluate the transport properties.
It is characteristics that the valence and conduction bands near the Fermi energy mainly consist of the $Tt$(Pb)-$p$ and $Ae$(Ba)-$d$ orbitals, respectively~\cite{SC_Sr3SnO}. In other words, a peculiar valence state $Tt^{4-}$ is realized here.
Another characteristic is the existence of the Dirac cones near the Fermi energy, which exhibits six-fold degeneracy in the Brillouin zone by crystal symmetry.

\begin{figure}
\begin{center}
\includegraphics[width=8.7 cm]{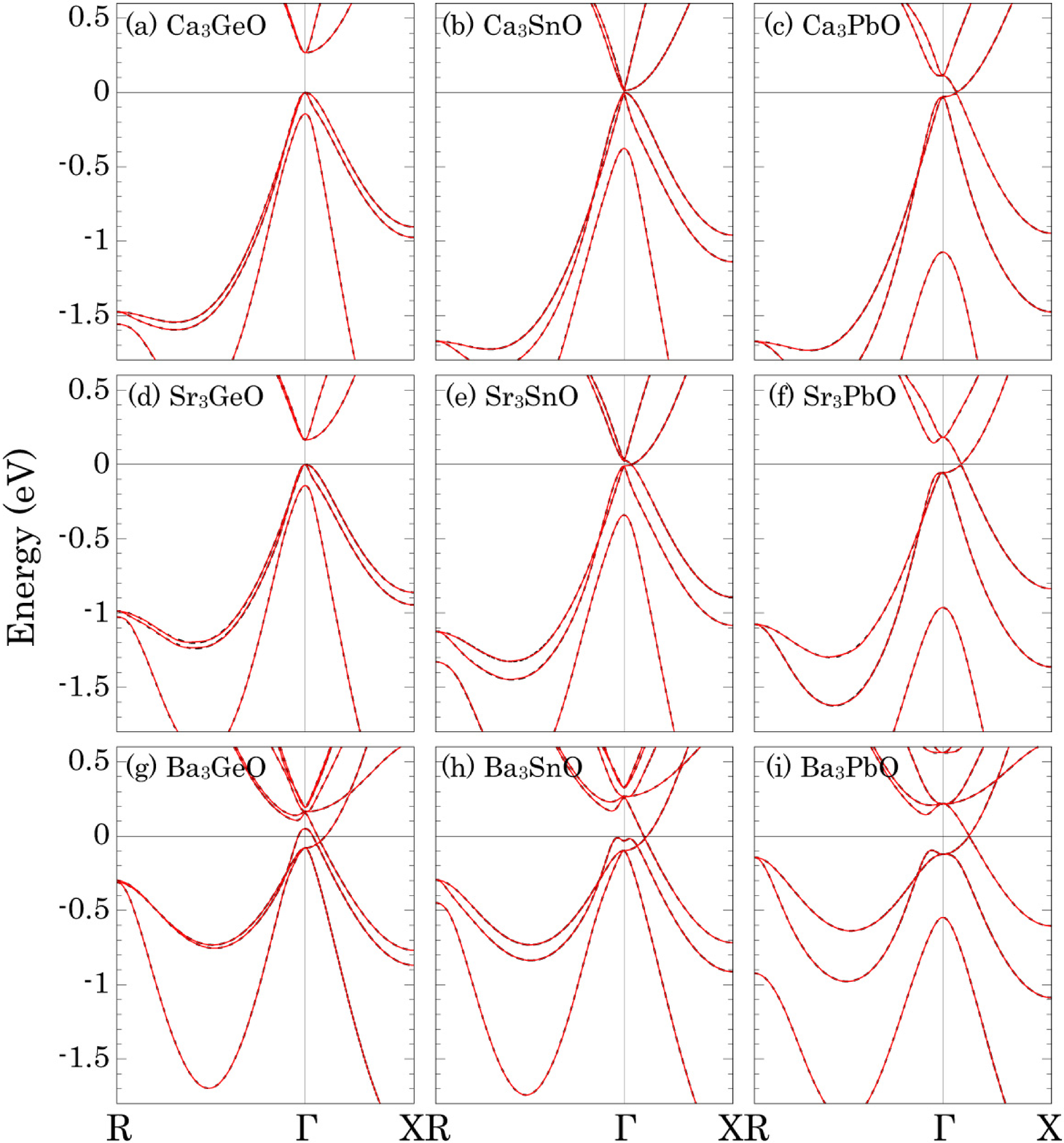}
\caption{Band structures of $Ae_3Tt$O ($Ae=$ Ca, Sr, Ba; $Tt=$ Ge, Sn, Pb) with the cubic structure (space group: $Pm\bar{3}m$). Black broken and red solid lines represent the band structures obtained with the first-principles calculation and the tight-binding model for the Wannier functions, respectively.}
\label{fig:band_o}
\end{center}
\end{figure}

Figure~\ref{fig:band_o} presents the band structures of various antiperovskite oxides, which shows that the existence of the band gap and the Dirac cones depends on the constituent elements. The atomic replacement of $Tt=$ Ge$\to$Sn$\to$Pb naturally leads to reduction of the band gap for $Ae=$ Ca and Sr, owing to an upward shift of the valence band dispersion mainly consisting of the $Tt$-atomic orbitals. This atomic replacement also enhances SOC, the effect of which shall be investigated later in this paper. Here, we only point out that the valence band splitting at the $\Gamma$ point induced by SOC becomes larger by this atomic replacement as shown in Fig.~\ref{fig:band_o}.

Quasi-one-dimensionality of the valence-top band structure of some materials is noteworthy.
For example, the valence-top band structure of Ca$_3$GeO shown in Fig.~\ref{fig:band_o}(a) consists of two nearly degenerate band dispersion with a heavy effective mass and the other one with a much lighter mass along the $\Gamma$-X line.
Concretely, the ratio of the effective masses along the $\Gamma$-X line for the top three (six when considering the spin degeneracy) valence bands is 3.1:0.49:1 for Ca$_3$GeO.
This feature corresponds with the fact that there are three quasi-one-dimensional band dispersion that are mobile along one of the $x$, $y$, and $z$ directions, respectively.
Such a quasi-one-dimensionality originates from the anisotropy of the $Tt$-$p$ orbitals.
Because low-dimensionality is desirable for high thermoelectric performance~\cite{Hicksone,Hicksone2,usuione,Fukuyama} owing to its large DOS near the band edge together with a sizable group velocity to a specific direction, the quasi-one-dimensionality in antiperovskites can be an advantageous feature for thermoelectric performance.

We note that, while each band dispersion has anisotropy along one of the $x$, $y$, and $z$ directions, the coexistence of these three bands result in the isotropy of the transport property, which is naturally expected for the cubic structure. Even though the transport is in total isotropic, the situation is much different from the case when two isotropic heavy bands coexist with one isotropic light band. Because for the latter case, the carrier with a large DOS has a heavy effective mass along all the directions, and so is not accompanied with the large group velocity. For more detailed investigation of this kind of band structure, i.e., coexisting low-dimensional bands with anisotropy along different directions, we refer the readers to Ref.~\onlinecite{Ochi_tetra}.
We also note that, in real one-dimensional systems, technological applications are not straightforward because they require high orientation of samples, without which the conductivity is easily lost. Low-dimensionality owing to the anisotropy of the electron wave function realized in rather isotropic crystal structure, like our target materials here, is favorable from this perspective~\cite{Ochi_tetra,Mori_SnSe,BiS2_thermo,BiS2_thermo_review} because the conductivity is expected to be kept without very high orientation of samples.

\subsubsection{Thermoelectric performance: comparison with experiment~\label{sec:TEO}}

Before proceeding to investigation on thermoelectric performance among antiperoviskite oxides,
we checked the consistency between our calculated values of the transport quantities and those reported in an experimental study.
In Ref.~\onlinecite{Okamoto}, $\rho=7.3$ m$\Omega$ cm and $S= 94$ $\mu$VK$^{-1}$ were reported for polycrystalline Ca$_3$SnO at 290 K.
The carrier density was also estimated from the Hall coefficients, $n=$ $1.44 \times 10^{19}$ and $1.43\times 10^{19}$ cm$^{-3}$ at 5 and 20 K, respectively. 
In our calculation for Ca$_3$SnO, we obtained $\rho \tau= 3.5\times 10^{-14}$ m$\Omega$ cm s and $S= 88$ $\mu$VK$^{-1}$ using the carrier density $n=1.43\times 10^{19}$ cm$^{-3}$ at 290 K.
The agreement with the experimental Seebeck coefficient (94 $\mu$VK$^{-1}$) is surprisingly good.
In addition, we can roughly estimate the relaxation time $\tau$ as $4.8\times 10^{-15}$ s, by taking a ratio of the calculated $\rho \tau$ and the experimental $\rho$. This is a typical length of the relaxation time for thermoelectric materials.

In the experimental study~\cite{Okamoto}, $\rho=2.5$ m$\Omega$ cm and $S= 22$ $\mu$VK$^{-1}$ at 290 K for polycrystalline Ca$_3$PbO were also reported.
For Ca$_3$PbO, we determined the carrier density so as to provide the Seebeck coefficient $S=22$ $\mu$VK$^{-1}$ at 290 K, which is the same value with the experimental Seebeck coefficient.
We obtained the carrier density $n=3.3\times 10^{20}$ cm$^{-3}$, and the calculated electrical resistivity for that carrier density is $\rho \tau=3.4\times 10^{-15} $ m$\Omega$ cm s. Therefore, by comparing it with the experimental $\rho$, the relaxation time $\tau$ was estimated as $1.3\times 10^{-15}$ s.
This length of the relaxation time is again typical for thermoelectric materials.
While it is shorter than the value assumed in this study, a longer relaxation time can be expected in future experiments because the sample investigated in Ref.~\onlinecite{Okamoto} is polycrystal.

\begin{figure}
\begin{center}
\includegraphics[width=8.4 cm]{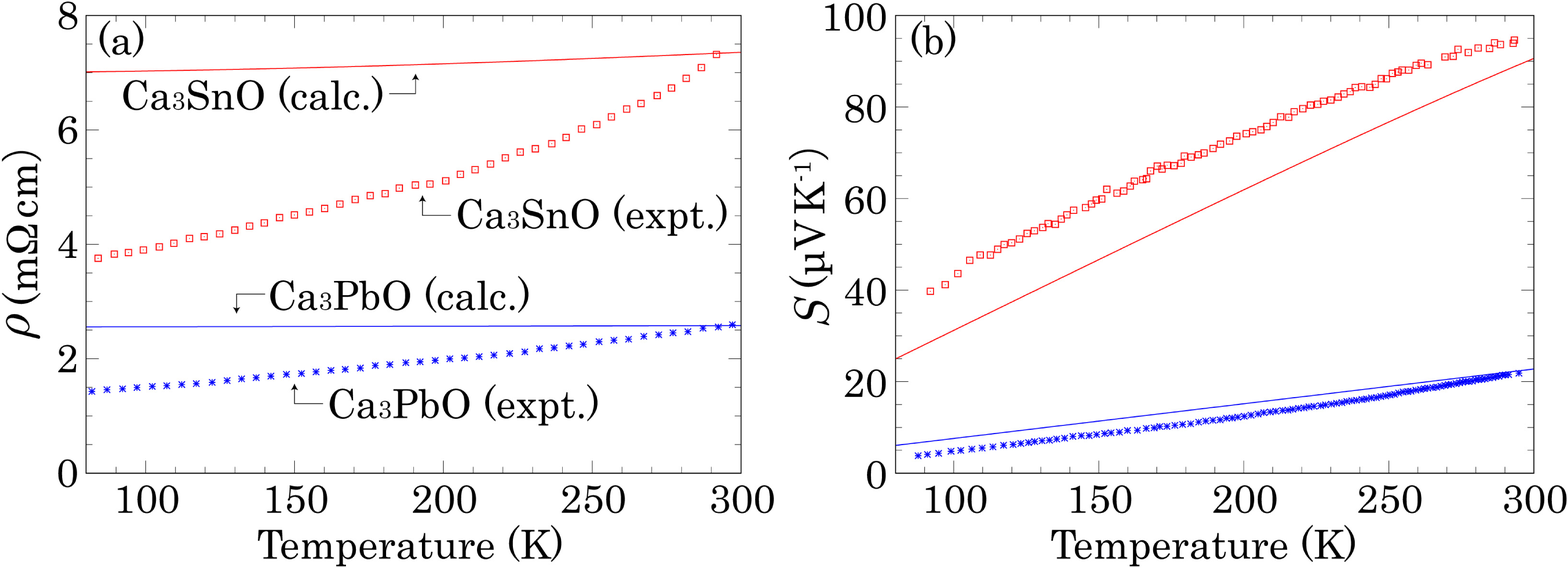}
\caption{(a) Electrical resistivity obtained by our calculation and experimental data taken from Ref.~\onlinecite{Okamoto} for Ca$_3$SnO and Ca$_3$PbO. (b) Those for the Seebeck coefficient. In calculation, we used the carrier density $n=1.43\times 10^{19}$ cm$^{-3}$ for Ca$_3$SnO and  $n=3.3\times 10^{20}$ cm$^{-3}$ for Ca$_3$PbO, and the constant relaxation time $\tau = 4.8\times 10^{-15}$ s for Ca$_3$SnO and $\tau = 1.3\times 10^{-15}$ s for Ca$_3$PbO, which were determined from the comparison with experiment and calculation as described in the main text.}
\label{fig:expt}
\end{center}
\end{figure}

By using the carrier density ($n=1.43\times 10^{19}$ cm$^{-3}$ for Ca$_3$SnO and  $n=3.3\times 10^{20}$ cm$^{-3}$ for Ca$_3$PbO)
and the relaxation time at 300 K ($\tau = 4.8\times 10^{-15}$ s for Ca$_3$SnO and $\tau = 1.3\times 10^{-15}$ s for Ca$_3$PbO) estimated above,
we can compare the temperature dependence of the electrical resistivity and the Seebeck coefficient between experiment and calculation.
Figure~\ref{fig:expt} presents the calculated electrical resistivity and the Seebeck coefficient and the experimental data taken from Ref.~\onlinecite{Okamoto}.
The calculated temperature dependence of the Seebeck coefficient shown in Fig.~\ref{fig:expt}(b) is to some extent consistent with the experimental one for these two materials. Possible origins for the error shown here are the error in determining the carrier density, the difference in the crystal structure, and
the accuracy of the calculated band structure.

A difference in the electrical resistivity between experiment and calculation shown in Fig.~\ref{fig:expt}(a) likely originates from the temperature dependence of the relaxation time, which was ignored in calculation.
For Ca$_3$SnO, the ratio of the electrical resistivity at 120 K and 290 K, $\rho_{\mathrm{290K}} \rho_{\mathrm{120K}}^{-1}$, is
1.0 in calculation using the constant relaxation-time approximation and the carrier density $n=1.4\times 10^{19}$ cm$^{-3}$, while $\rho_{\mathrm{290K}} \rho_{\mathrm{120K}}^{-1}=1.8$ in experiment~\cite{Okamoto}. This difference suggests that $\tau_{\mathrm{120K}} \tau_{\mathrm{290K}}^{-1}$ is about 1.8 $(=1.8/1.0)$.
On the other hand, the thermal conductivity of Ca$_3$SnO reported in Ref.~\onlinecite{Okamoto}, which can be regarded as $\kappa_{\mathrm{lat}}$ because the electronic contribution of the thermal conductivity was reported to be negligible, is around 2.9 and 1.7 Wm$^{-1}$K$^{-1}$ at 120 and 290 K, respectively.
Thus, $\kappa_{\mathrm{lat,120K}} \kappa_{\mathrm{lat,290K}}^{-1}$ is about 1.7.
Here, the calculated $ZT$ within the constant relaxation-time approximation depends only on the ratio $\kappa_{\mathrm{lat}}\tau^{-1}$ because
\begin{equation}
ZT = \frac{\sigma S^2 T}{\kappa_{\mathrm{el}}+\kappa_{\mathrm{lat}}} =  \frac{\sigma S^2 T\tau^{-1}}{\kappa_{\mathrm{el}}\tau^{-1}+\kappa_{\mathrm{lat}}\tau^{-1}},
\end{equation}
where both $\sigma S^2 T\tau^{-1}$ and $\kappa_{\mathrm{el}}\tau^{-1}$ depend neither on $\tau$ nor on $\kappa_{\mathrm{lat}}$.
Since $\kappa_{\mathrm{lat}}\tau^{-1}$ is expected to be roughly the same between $T=120$ and 290 K as shown here,
we can expect that the calculated temperature dependence of $ZT$ is to some extent reliable.
We note that the temperature dependence of $\tau$ and $\kappa_{\mathrm{lat}}$ in the high-temperature region is at present not available in experiment.
Therefore, the calculated temperature dependence of $ZT$ beyond room temperature still has some uncertainty regarding these assumed values of $\tau$ and $\kappa_{\mathrm{lat}}$.

\subsubsection{Thermoelectric performance: calculation}
\begin{figure}
\begin{center}
\includegraphics[width=8.4 cm]{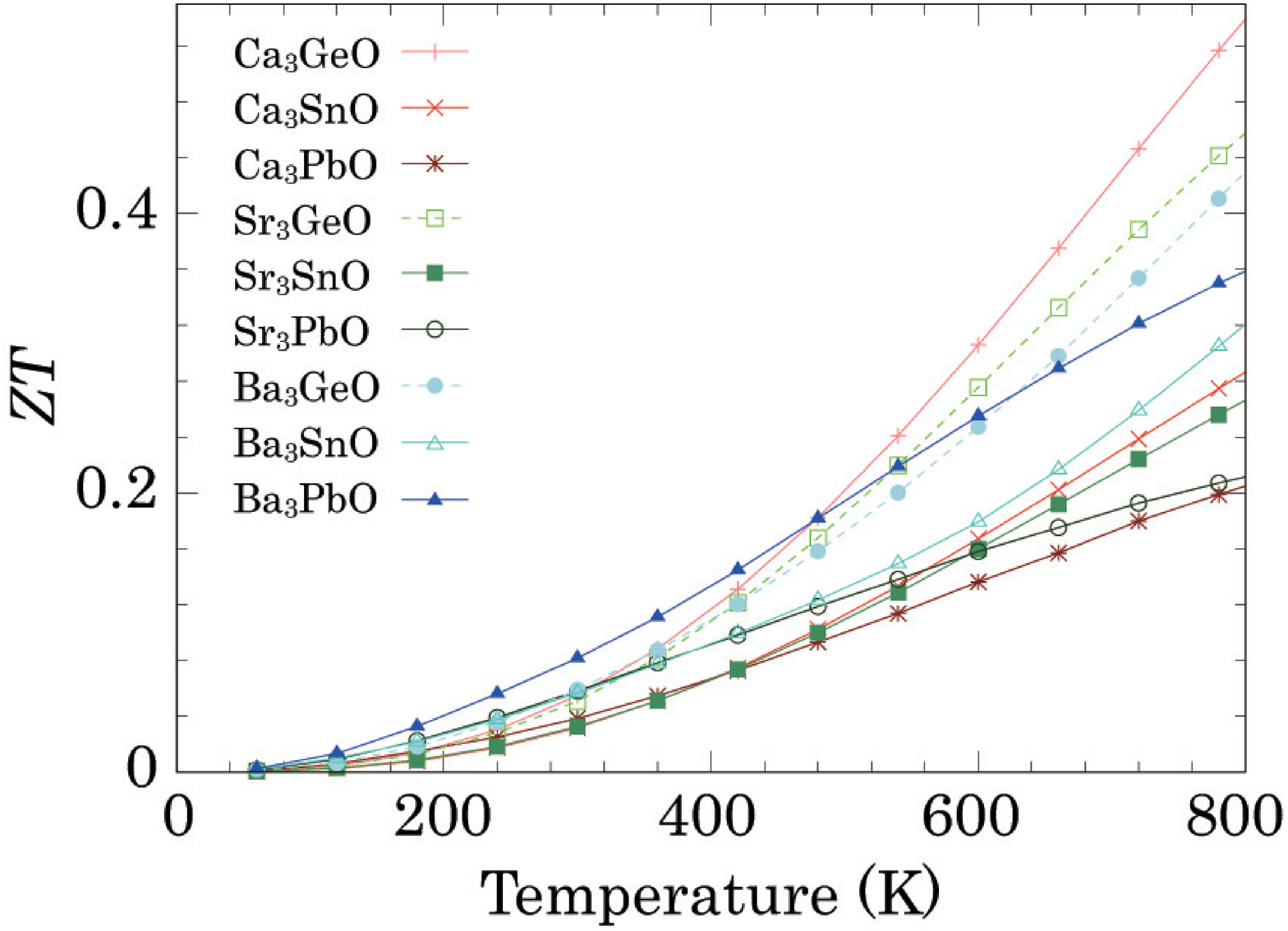}
\caption{Calculated $ZT$ values of $Ae_3Tt$O ($Ae=$ Ca, Sr, Ba; $Tt=$ Ge, Sn, Pb) with the cubic structure (space group: $Pm\bar{3}m$) with respect to temperature. Broken lines correspond to materials that have yet to be synthesized in the structure with the $Pm\bar{3}m$ space group. The hole carrier concentration was optimized for each point.}
\label{fig:ZT_o}
\end{center}
\end{figure}

Figure~\ref{fig:ZT_o} presents the calculated $ZT$ values of the antiperovskite oxides with the cubic ($Pm\bar{3}m$) structure.
Broken lines in the figure correspond to materials that have yet to be synthesized in the structure with the $Pm\bar{3}m$ space group.
The hole carrier concentration was optimized for each point, and so depends on temperature in this plot.

In the high-temperature region, Ca$_3$GeO yields the highest $ZT$, where the quasi-one-dimensional band structure together with a finite band gap is realized as shown in Fig.~\ref{fig:band_o}(a). This result seems to be natural because of the superiority of the low-dimensional electronic structure for thermoelectric performance as described in the previous section. In addition, the band gap prevents cancellation of the contribution from the electron and hole carriers in the transport coefficient $K_1$, Eq.~(\ref{eq:transp}), appearing in the Seebeck coefficient, Eq.~(\ref{eq:sigmas}).

On the other hand, in the low-temperature region, it is rather counterintuitive that Ba$_3$PbO yields the highest $ZT$.
In its band structure shown in Fig.~\ref{fig:band_o}(i), there are Dirac cones at the Fermi energy without the gap.
As mentioned in the previous paragraph, the band dispersion without the gap is generally not favorable for thermoelectric performance because of the cancellation of the electron- and hole-carrier transport.
In the next section, we investigate the way how the Dirac cones in Ba$_3$PbO result in the high thermoelectric performance.

\subsubsection{How do the Dirac cones in Ba$_3$PbO enhance the thermoelectric performance?}

\begin{figure}
\begin{center}
\includegraphics[width=6 cm]{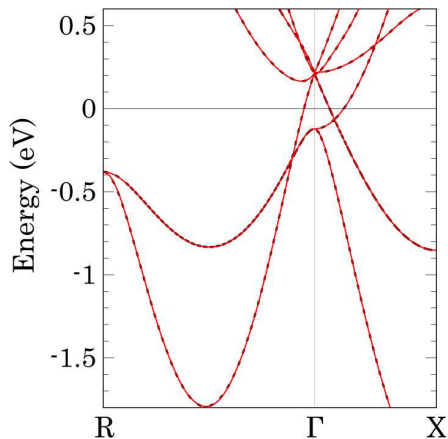}
\caption{Band structures of Ba$_3$PbO with the cubic structure (space group: $Pm\bar{3}m$). Calculation was performed without the inclusion of SOC. Black broken and red solid lines represent the band structures obtained with the first-principles calculation and the tight-binding model for the Wannier functions, respectively.}
\label{fig:ba3pbo_woso}
\end{center}
\end{figure}

To begin with, we investigated the effect of SOC on the thermoelectric performance.
Figure~\ref{fig:ba3pbo_woso} presents the calculated band structure of Ba$_3$PbO without the inclusion of SOC.
Although the Dirac cones are preserved along the $\Gamma$-X line, the system becomes metallic (i.e. a finite-size Fermi surface takes place) without SOC as seen in the band dispersion along the $\Gamma$-R line.
Metallic electronic structure is clearly unfavorable for thermoelectric performance.
In fact, the maximum value of $ZT$ with respect to the hole carrier concentration is 0.04 for $T=300$ K when SOC is switched off, which is only half of $ZT=0.08$ when SOC is included in calculation.
Therefore, SOC is one of the indispensable factors for high thermoelectric performance of Ba$_3$PbO.

\begin{figure}
\begin{center}
\includegraphics[width=8.0 cm]{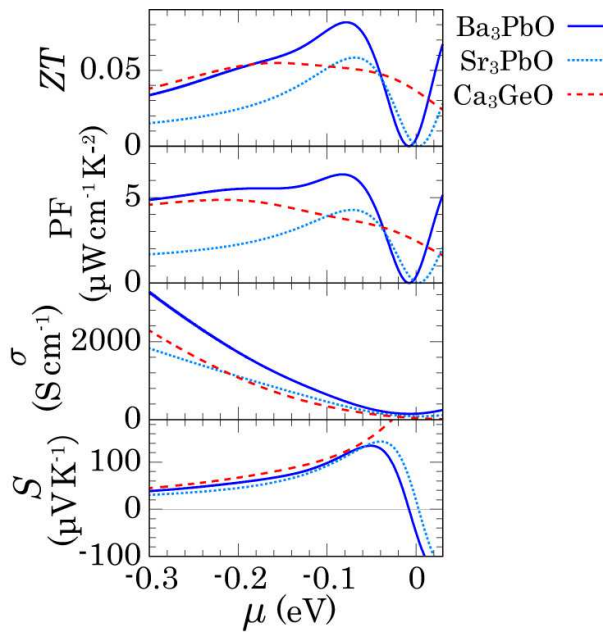}
\caption{Calculated $ZT$, PF, $\sigma$, and $S$ at 300 K for Ba$_3$PbO, Sr$_3$PbO, and Ca$_3$GeO with the cubic structure (space group: $Pm\bar{3}m$). The Fermi energy without doped carriers was set to zero.}
\label{fig:3mat}
\end{center}
\end{figure}

To obtain more insight, we compared several transport quantities for three cases: Ba$_3$PbO, Sr$_3$PbO, and Ca$_3$GeO.
From here on, we again included SOC in the calculations.
Among these three compounds, Ba$_3$PbO and Sr$_3$PbO have Dirac cones while Ca$_3$GeO is an insulator with a gap as shown in Fig.~\ref{fig:band_o}(a)(f)(i).
The Dirac cones without the gap allows the cancellation of the electron and hole carrier contribution for the Seebeck coefficient $S$ when the carrier concentration is low, which was verified by our calculation results at 300 K as shown in Fig.~\ref{fig:3mat}.
However, we also notice that the Seebeck coefficient becomes comparable for these three materials when the chemical potential is sufficiently far from $\mu=0$, and $ZT$ and PF reach their maximum values in such a region.
In fact, even for Ba$_3$PbO and Sr$_3$PbO, the Seebeck coefficient can exceed 100 $\mu$VK$^{-1}$.
This is one of the reasons why the relatively large values of $ZT$ and PF can be achieved for the non-gap band dispersion for these materials.
We note that, at high temperatures, the cancellation of the electron and hole carrier contribution occurs for wider carrier concentration, and so $ZT$ of Ba$_3$PbO with the non-gapped Dirac cones is in fact much lower than that of Ca$_3$GeO with the gapped band structure as we have seen in Fig.~\ref{fig:ZT_o}.

By looking into the electrical conductivity $\sigma$ and the Seebeck coefficient $S$ shown in Fig.~\ref{fig:3mat}, a superiority of Ba$_3$PbO among the three materials can be seen for its high electrical conductivity.
A key characteristic of its band structure around $\mu = -0.08$ eV, where $ZT$ and PF are maximized, is the high valley degeneracy. 
First, the Dirac cones have six-fold degeneracy by the crystal symmetry.
In addition, the valence-top band structure around the $\Gamma$ and R points can enhance the thermoelectric performance by temperature broadening (see Fig.~\ref{fig:band_o}(i)). 
On the other hand, the band structure without the Dirac cones as in Ca$_3$GeO has no valley degeneracy since there is no other $k$-points that are equivalent to the $\Gamma$ point where the valence-band top resides.
We note that, in Sr$_3$PbO, the R valley is too deep to enhance the thermoelectric performance while the $\Gamma$ point can play a role, which is an important difference between Sr$_3$PbO and Ba$_3$PbO.
It has been established that the multi-valley band structure is favorable for high thermoelectric performance~\cite{band_conv_PbCh}.
Therefore, the high valley degeneracy is an important advantage of Ba$_3$PbO.
It is also interesting that the valence-band top at the $\Gamma$ point for Ba$_3$PbO has a pudding-mold-shape~\cite{pudding}, which can enhance DOS near the band edge. This is another outcome of the large band deformation near the Fermi energy induced by SOC.

Before proceeding to the next section, we point out two issues regarding thermoelectric performance of the Dirac cone.
First, one should pay attention to the applicability of the Boltzmann transport theory with the constant-relaxation-time approximation because it cannot appropriately deal with the inter-band scattering effects, which can affect thermoelectric properties of the systems where the bipolar effects are important such as those possessing the Dirac cone~\cite{Yamamoto_bip}.
In our case, the chemical potential becomes sufficiently deep ($\sim$ $-0.1$ eV) from the Dirac points at room temperature.
Therefore, we expect that the inter-band scattering is not so dominant for our calculated results compared with the case when the chemical potential lies near the Dirac point, at least if the scattering strength is not so strong that the inter-band scattering becomes very active.
Second, we point out that a possible long relaxation time is another advantage of the Dirac cones.
This feature is naturally expected because a small DOS such as for the Dirac cone generally reduces the number of possible electron scattering processes and then yields a long relaxation time. However, these points regarding the scattering processes need further investigation and so are important future issues.

\subsection{Nitrides $Ae_3Pn$N with the cubic structure~\label{sec:n}}

Next, we move on to the antiperovskite nitrides with a chemical formula $Ae_3Pn$N ($Ae=$ Ca, Sr, Ba; $Pn=$ As, Sb, Bi). In the same manner as the previous section, we assumed the cubic structure with a space group $Pm\bar{3}m$ for all the compounds in this section. 
We shall see thermoelectric performance of these materials with other crystal symmetries in Sec.~\ref{sec:struct}.

\subsubsection{Band structures}

\begin{figure}
\begin{center}
\includegraphics[width=8.5 cm]{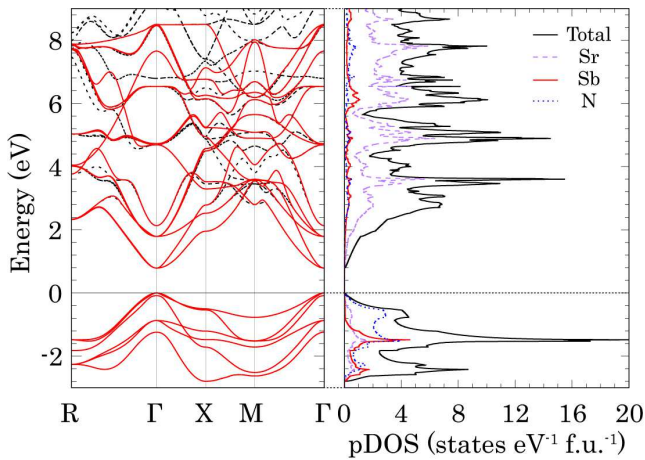}
\caption{Band structure and pDOS of Sr$_3$SbN with the cubic structure (space group: $Pm\bar{3}m$). Black broken and red solid lines in the band structure represent the band structures obtained with the first-principles calculation and the tight-binding model for the Wannier functions, respectively.}
\label{fig:band_sr3sbn}
\end{center}
\end{figure}

Figure~\ref{fig:band_sr3sbn} presents the whole band structure and pDOS of Sr$_3$SbN as a typical member of antiperovskite nitrides $Ae_3Pn$N.
The most striking difference from oxides is that the valence-top band structure mainly consists of nitrogens, which holds also for other nitrides investigated in this study.
This is because nitrogen atomic orbitals have shallower energy levels than oxygen atomic orbitals.
In addition, $Pn$ atomic orbitals have deeper energy levels than $Tt$ atomic orbitals in the same period.

\begin{figure}
\begin{center}
\includegraphics[width=8.7 cm]{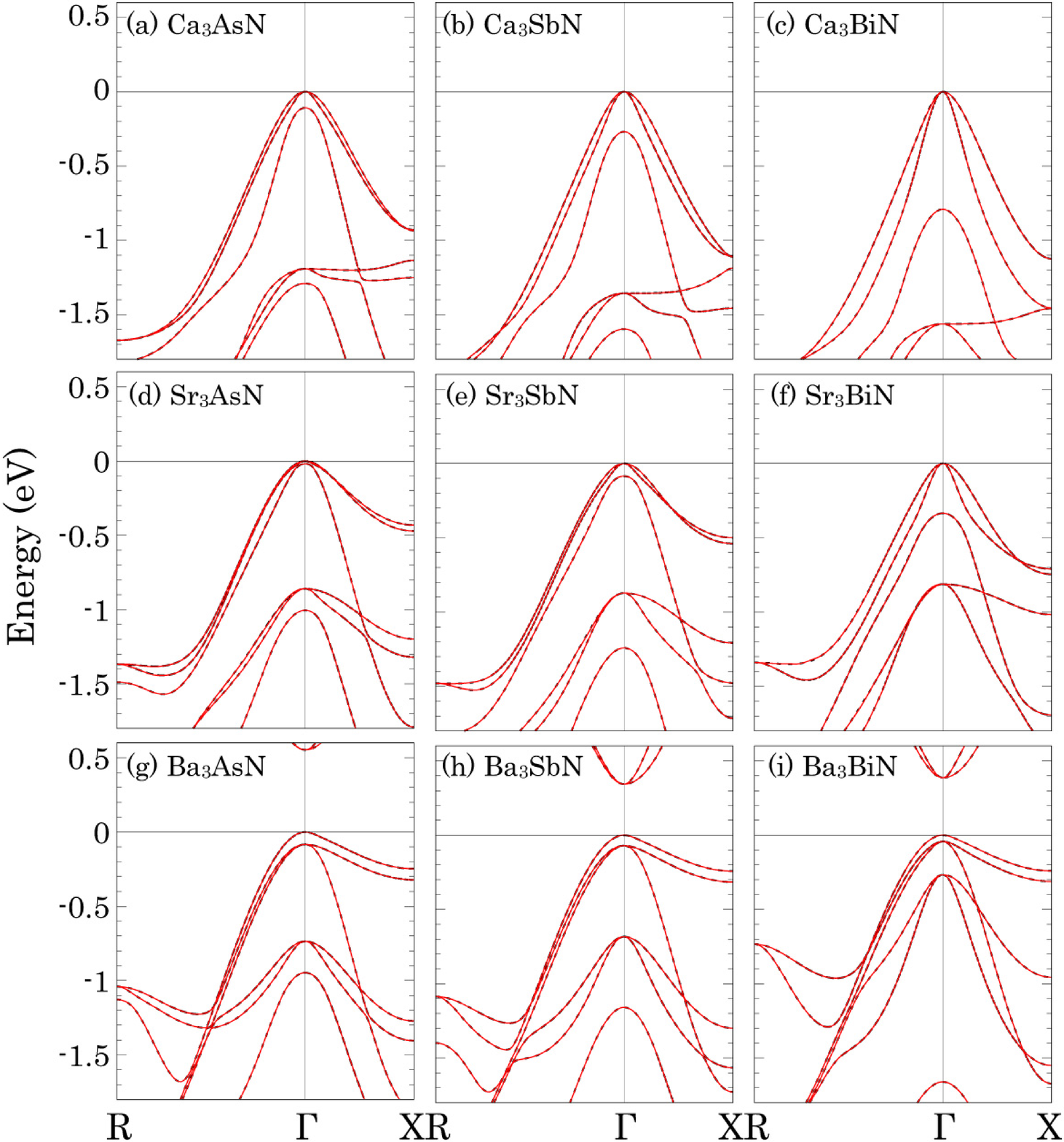}
\caption{Band structures of $Ae_3Pn$N ($Ae=$ Ca, Sr, Ba; $Pn=$ As, Sb, Bi) with the cubic structure (space group: $Pm\bar{3}m$). Black broken and red solid lines represent the band structures obtained with the first-principles calculation and the tight-binding model for the Wannier functions, respectively.}
\label{fig:band_n}
\end{center}
\end{figure}

As a result, the band structures of the antiperovskite nitrides shown in Fig.~\ref{fig:band_n} have different features from oxides.
First of all, unlike the oxides, all the nitrides have a band gap.
Therefore, there is no chance that the Dirac cones appear at the Fermi energy.
On the other hand, we can recognize the low dimensionality of the valence-top band structure similar to that seen in Fig.~\ref{fig:band_o}(a).
This observation corresponds to the fact that the three $p$ orbitals of nitrogens have anisotropy (quasi-one-dimensionality) for its conduction.
We shall see this point in more detail in the next section.

\subsubsection{Thermoelectric performance~\label{sec:TPF_n}}

\begin{figure}
\begin{center}
\includegraphics[width=8.4 cm]{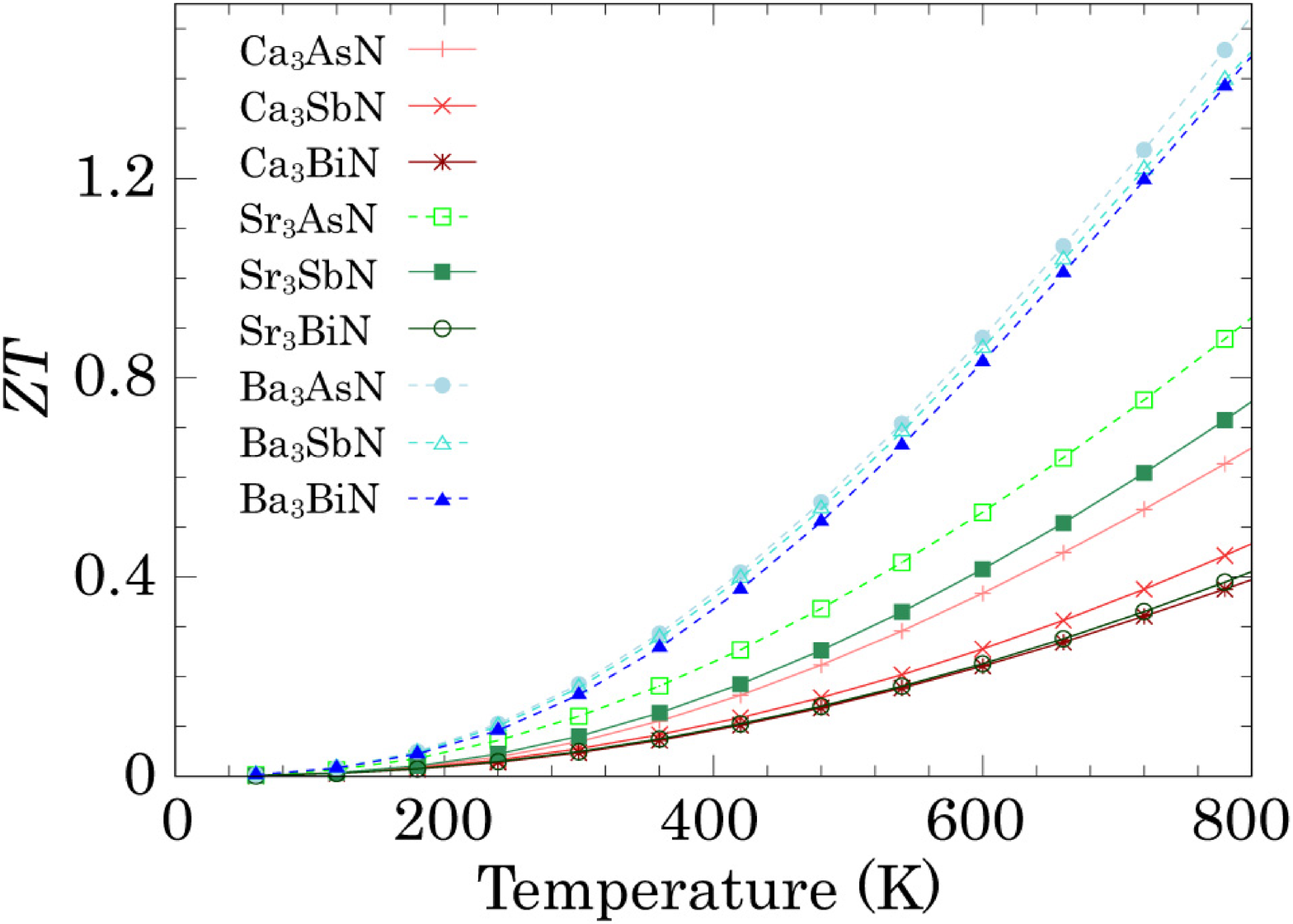}
\caption{Calculated $ZT$ values of $Ae_3Pn$N ($Ae=$ Ca, Sr, Ba; $Pn=$ As, Sb, Bi) with the cubic structure (space group: $Pm\bar{3}m$) with respect to temperature. Broken lines correspond to materials that have yet to be synthesized in the structure with the $Pm\bar{3}m$ space group. The hole carrier concentration was optimized for each point.}
\label{fig:ZT_n}
\end{center}
\end{figure}

Calculated $ZT$ values of the antiperovskite nitrides are shown in Fig.~\ref{fig:ZT_n}.
Because of the low-dimensionality and a sufficiently large band gap, calculated $ZT$ for nitrides are relatively high.
However, we note that, materials with high $ZT$ values are not stable as the cubic structure.
If one restricts the target materials to those existing as the cubic structure in experiment, which are shown with solid lines in Fig.~\ref{fig:ZT_n},
the maximum $ZT$ value at 300 K is 0.08 for Sr$_3$SbN, which is comparable to that of Ba$_3$PbO, 0.08.
Even under this restriction, at high temperatures, the estimated $ZT$ values of some nitrides such as Sr$_3$SbN and Ca$_3$AsN exceed those of all the oxides calculated in this study.

\begin{figure}
\begin{center}
\includegraphics[width=7 cm]{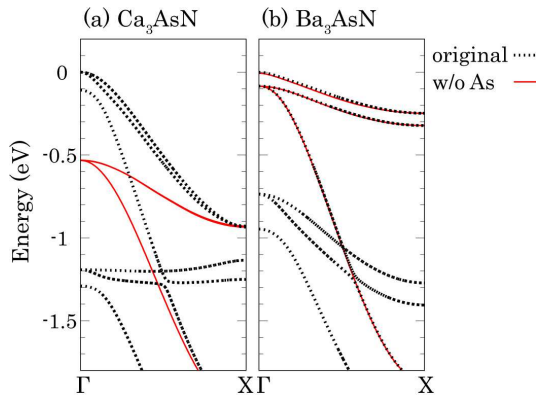}
\caption{Band structures calculated using the tight-binding model for (a) Ca$_3$AsN and (b) Ba$_3$AsN. Black dotted and red solid lines represent the original band structure and that calculated without the As orbitals, respectively.}
\label{fig:model_n}
\end{center}
\end{figure}

Why do the crystal structures that are unstable as the cubic structure exhibit high $ZT$ values in calculation?
A key is the correlation between the strong quasi-one-dimensionality of the band structures of Sr$_3$AsN and Ba$_3Pn$N ($Pn=$ As, Sb, Bi) and their high $ZT$ values shown in Fig.~\ref{fig:ZT_n}.
In other words, as shown in Fig.~\ref{fig:band_n}(d)(g)(h)(i), the ratio of the effective masses of the heavy and light band dispersions along the $\Gamma$-X line looks large for these compounds with high $ZT$ values.

To see why these materials exhibit the strong quasi-one-dimensionality, we calculated band structures of Ca$_3$AsN and Ba$_3$AsN using our tight-binding model from which the As orbitals were excluded. In other words, the tight-binding Hamiltonian consisting only of the $Ae$ and N orbitals was solved here. Obtained band structures are shown with red solid lines in Fig.~\ref{fig:model_n} and compared with the original band structures shown with black dotted lines.
While the valence-top band structure of Ba$_3$AsN shown in Fig.~\ref{fig:model_n}(b) is almost unaffected by neglecting the As orbitals,
we found that the low-dimensionality of the nitrogen bands in Ca$_3$AsN is much degraded by hybridization with the As orbitals as shown in Fig.~\ref{fig:model_n}(a).
Concretely, the ratios of the effective masses along the $\Gamma$-X line for the top three (six when considering the spin degeneracy) valence bands in Fig.~\ref{fig:model_n} are
1.5:0.64:1 for the original Ca$_3$AsN, 1.3:1.3:1 for the Ca$_3$AsN without the Wannier orbitals of As,
1.7:3.4:1 for the original Ba$_3$AsN, and 1.6:3.6:1 for the Ba$_3$AsN without the Wannier orbitals of As.
This difference can be naturally understood because, in Ba$_3$AsN, the Ba ionic radius may be too large to keep the As-N distance short enough to hybridize, which preserves the low-dimensionality of the nitrogen orbitals.
Therefore, this strong low-dimensionality is in accord with the instability of the crystal structure.
In general, the structural instability might enhance the anharmonicity of phonons, which often reduces the lattice thermal conductivity and then enhances $ZT$. It is interesting that the structural instability and the improvement of the electronic band structure, both of which might be favorable for $ZT$, can occur simultaneously in the antiperovskite nitrides.
We note that this expectation should be checked carefully because there are other factors that involve with $ZT$ such as a change in the electron relaxation time by increasing the structural instability. First-principles evaluation of these quantities is an interesting future issue.

\subsection{Thermoelectric performance with the orthorhombic and hexagonal structures~\label{sec:struct}}

In this section, we investigated thermoelectric performance of the antiperovskite oxides and nitrides with the orthorhombic and hexagonal structures.
For this purpose, we first evaluated which structure among the experimentally observed space groups is the most stable for each nitride because some nitrides have yet to be synthesized in experiment, unlike the oxides.
After that, we discussed the effect of the structural change from the cubic structure on thermoelectric performance.
Temperature dependence of $ZT$ for the most promising candidates we found are also presented.

\subsubsection{Stability of the $Pm\bar{3}m$ phase~\label{sec:stab}}

\begin{figure}
\begin{center}
\includegraphics[width=8.5 cm]{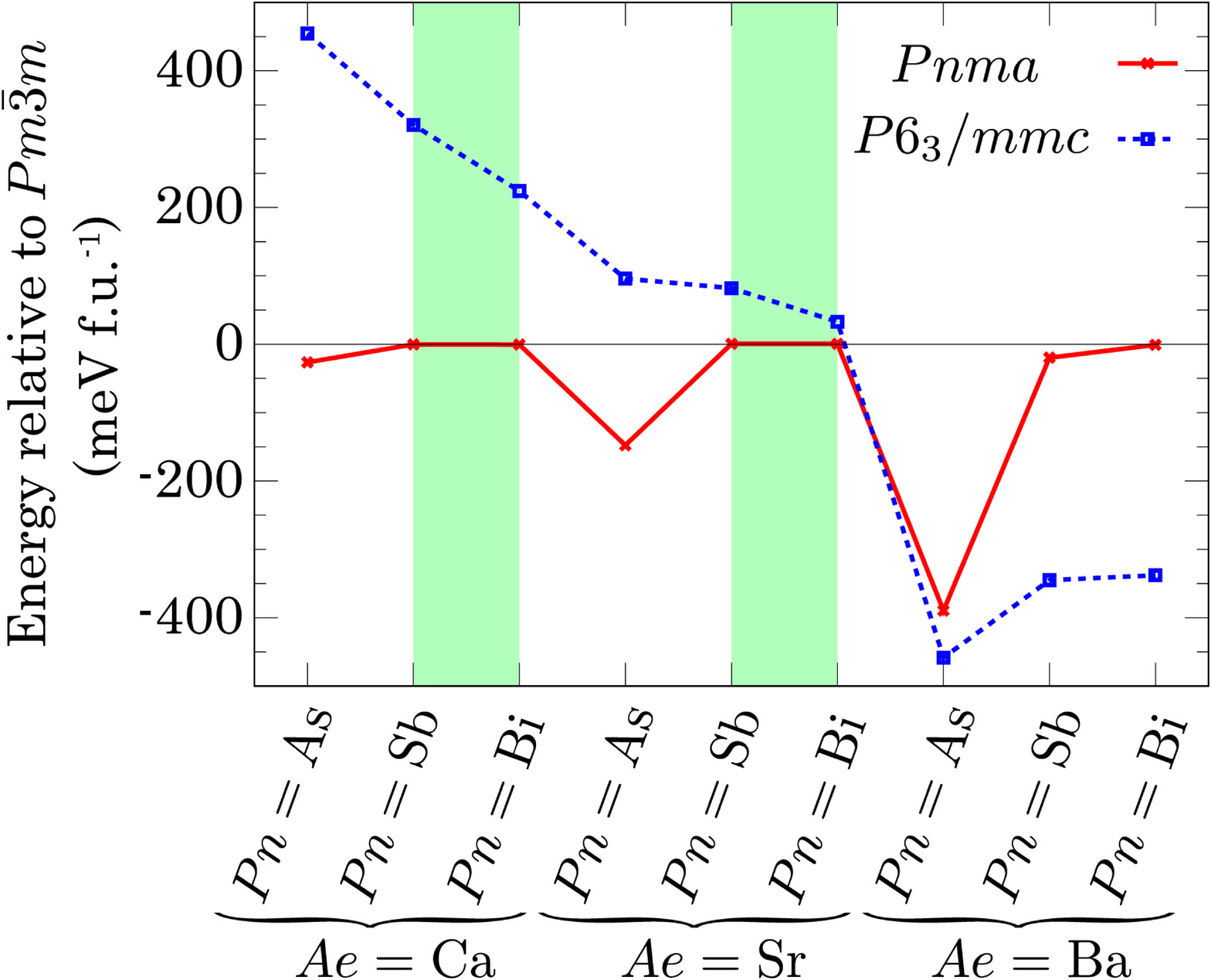}
\caption{Total energies of the $Pnma$ and $P$6$_3$/$mmc$ phases relative to that of the $Pm\bar{3}m$ for $Ae_3Pn$N ($Ae=$ Ca, Sr, Ba; $Pn=$ As, Sb, Bi). Materials where $Pm\bar{3}m$ is the most stable in our calculation (see the main text for more detail) are shaded by color.}
\label{fig:stab}
\end{center}
\end{figure}

We investigated the structural stability of the $Pm\bar{3}m$ phase for the nitrides,
by comparing its total energy with that of the $Pnma$ and $P$6$_3$/$mmc$ phases.
Figure~\ref{fig:stab} presents the total energies of the $Pnma$ and $P$6$_3$/$mmc$ phases, relative to that of the $Pm\bar{3}m$ phase, for the antiperovskite nitrides.
Because $Pnma$ is a subgroup of $Pm\bar{3}m$, the zero relative energy of the former to the latter means that the crystal structure becomes $Pm\bar{3}m$  in calculation even when one allows the crystal distortion that can take place for the $Pnma$ space group.
To be more precise, when the total energy difference of these two phases is less than 0.1 meV f.u.$^{-1}$, we regarded the crystal structure of the $Pnma$ phase falls into the $Pm\bar{3}m$ phase. Materials with the $Pm\bar{3}m$ phase as the most stable structure are shaded by color in the figure.
We note that, because $Pnma$ is a subgroup of $Imma$, we need not to calculate the total energy of the $Imma$ phase for discussing the stability of the cubic structure. In addition, we have verified that the optimized crystal structures for $Ae$AsN$_3$ and BaSbN$_3$ no longer have the full $Imma$ symmetry, i.e., the crystal structure seems to fall into the $Pnma$ phase for these compounds.

\begin{figure}
\begin{center}
\includegraphics[width=8.5 cm]{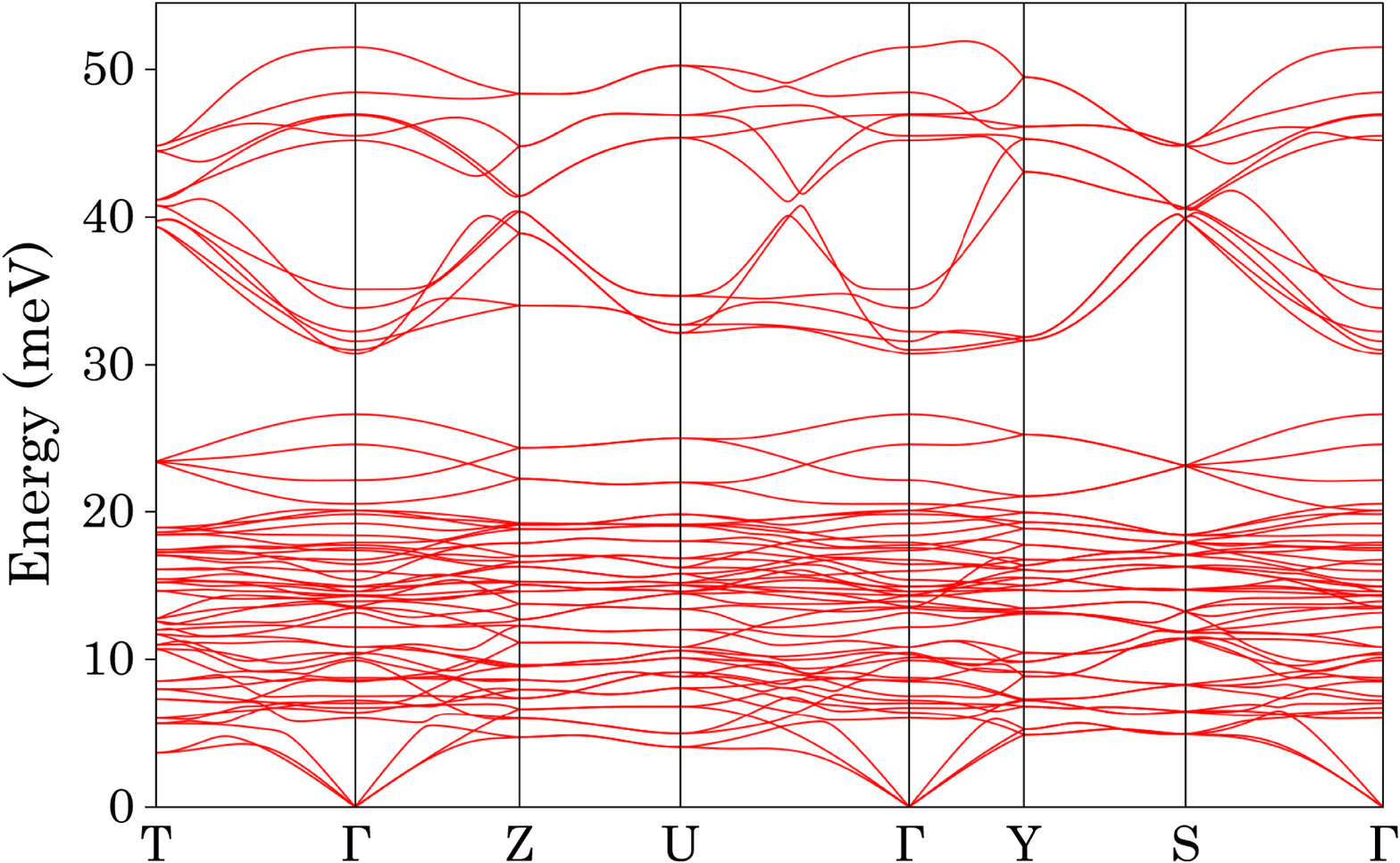}
\caption{Phonon dispersion of Sr$_3$AsN in the $Pnma$ phase. The $\bm{k}$-points represented with the crystal coordinate are as follows: T $= ({\bm b^*}+{\bm c^*})/2$, Z $={\bm c^*}/2$, U $=({\bm a^*}+{\bm c^*})/2$, Y $ = {\bm b^*}/2$, and S $= ({\bm a^*}+{\bm b^*})/2$.}
\label{fig:ph}
\end{center}
\end{figure}

Our results presented in Fig.~\ref{fig:stab} show surprisingly good agreement with the experimental observation listed in Table~\ref{table:crys_n}, with respect to the most stable structure of each compound.
In addition, the experimental observation that Ca$_3$AsN becomes $Pm\bar{3}m$ at high temperature is also consistent with a small energy difference between the $Pnma$ and $Pm\bar{3}m$ phases in our calculation.
It is characteristic that the $Pm\bar{3}m$ phase in all the $Ae=$ Ba compounds is much unstable than the $P$6$_3$/$mmc$ phase.
We found that the unsynthesized Sr$_3$AsN and Ba$_3$AsN are likely to be the $Pnma$ and $P$6$_3$/$mmc$ phases, respectively.
Because Sr$_3$AsN with the $Pnma$ structure is a good candidate for high-performance thermoelectric material as we shall see later in this paper, we also performed phonon calculation for it to verify the stability of this structure. For this purpose, we employed the finite displacement method as implemented in the \textsc{Phonopy}\cite{phonopy} software in combination with VASP.
We used a $2\times 2\times 2$ $\bm{q}$-mesh without including the spin-orbit coupling because Sr$_3$AsN consists of relatively light elements. We used a $4\times 3\times 4$ $\bm{k}$-mesh for calculation of the corresponding ($2\times 2\times 2$) supercell. Figure~\ref{fig:ph} presents the calculated phonon dispersion of Sr$_3$AsN with the $Pnma$ structure. Because no imaginary modes appear here, we can conclude that this structure is dynamically stable~\cite{note_phonon}.

We next investigated thermoelectric performance of materials with the hexagonal and orthorhombic structures.
On the basis of the above calculation results shown in Fig.~\ref{fig:stab} for nitrides and the experimental observation shown in Table~\ref{table:crys_o} for oxides, we focused on materials that are stable in these structures rather than as the cubic phase.

\subsubsection{Hexagonal structure}

\begin{figure}
\begin{center}
\includegraphics[width=5 cm]{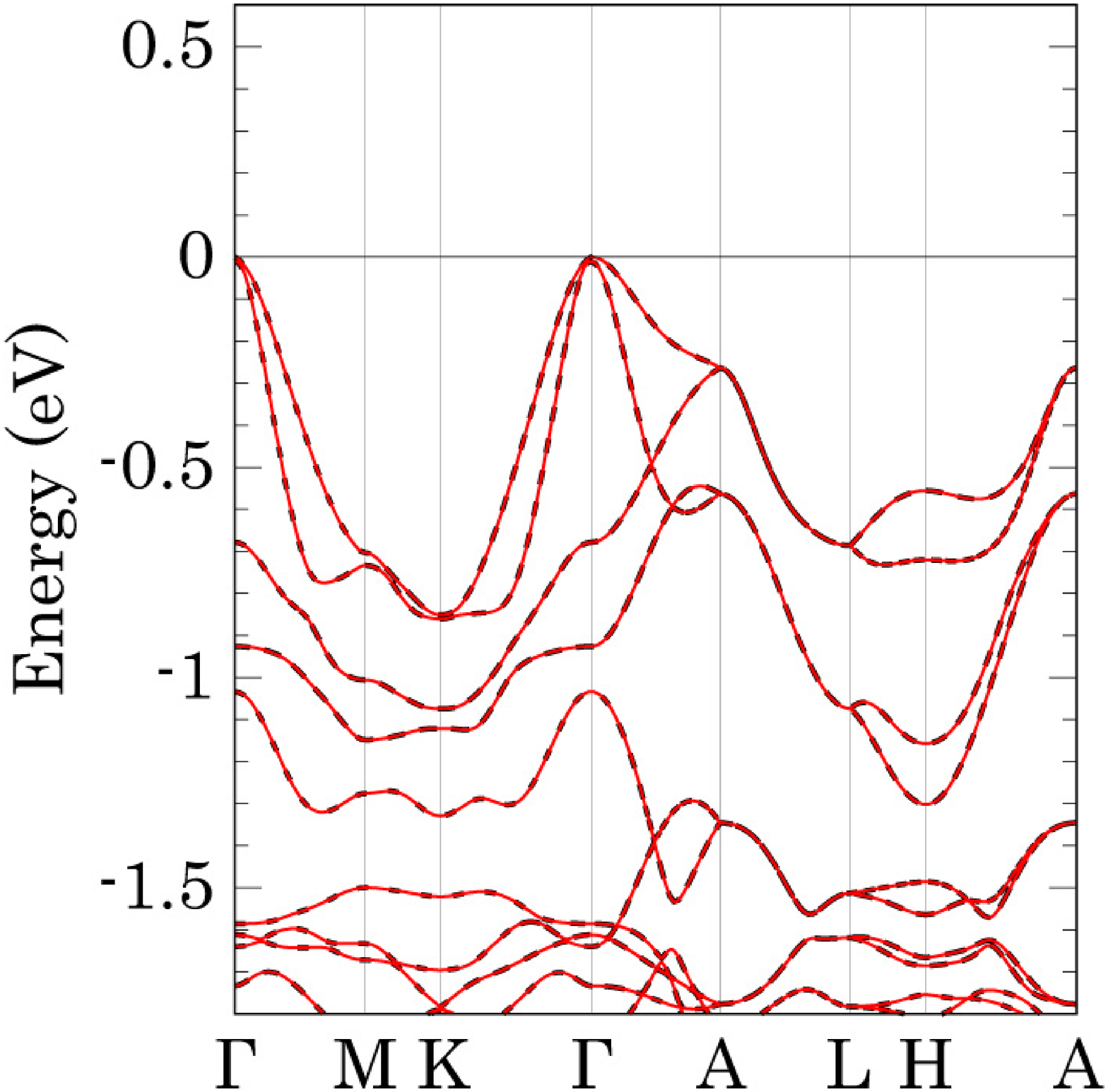}
\caption{Band structures of Ba$_3$BiN with the hexagonal structure (space group: $P$6$_3$/$mmc$). Black broken and red solid lines represent the band structures obtained with the first-principles calculation and the tight-binding model for the Wannier functions, respectively.}
\label{fig:band_hexa}
\end{center}
\end{figure}

\begin{table}
\caption{\label{table:ZT_hexa} Calculated $ZT$ values of cubic and hexagonal Ba$_3Pn$N at 300 K. Hole carrier concentration was optimized for each condition.}
\begin{center}
\begin{tabular}{c c c c}
\hline \hline
lattice & cubic& \multicolumn{2}{c}{hexagonal}\\
axis & $a=b=c$ & $a=b$ & $c$ \\
\hline
Ba$_3$AsN & 0.18 & 0.11 & 0.11\\
Ba$_3$SbN & 0.18 & 0.06 & 0.10\\
Ba$_3$BiN & 0.16 & 0.07 & 0.05\\
\hline \hline
\end{tabular}
\end{center}
\end{table}

Figure~\ref{fig:band_hexa} presents the band structure of Ba$_3$BiN with the hexagonal structure (space group: $P$6$_3$/$mmc$). 
Compared with that for the cubic structure as shown in Fig.~\ref{fig:band_n}(i), this band structure does not seem to be favorable for thermoelectric performance because of the weakened anisotropy of the band dispersion near the valence-band top. As a matter of fact, the maximum value of $ZT$ at 300 K is around 0.07 for the $a$ and $b$ directions and 0.05 for the $c$ direction, which are less than a half of the $ZT$ value for the cubic structure, 0.16.
As shown in Table~\ref{table:ZT_hexa}, the situation is similar to Ba$_3$AsN and Ba$_3$SbN, for which the hexagonal structure was the most stable in our calculation presented in the previous section.

Whereas the electronic band structure is unfavorable for thermoelectric performance, a possible rattling motion~\cite{clathrates,rattling1,rattling2,rattling3,rattling4,rattling5,rattling6,rattling7,rattling8} of the $Pn$ atoms, which can reduce the lattice thermal conductivity and thus increase $ZT$, is intriguing in the hexagonal phase~\cite{APSmarch}.

\subsubsection{Orthorhombic structures~\label{sec:strct_o}}

\begin{table}
\caption{\label{table:ZT_ortho1} Calculated $ZT$ values of the cubic ($Pm\bar{3}m$) and orthorhombic ($Imma$) structures at 300 K. Calculation results are only shown for materials for which the $Imma$ structure was experimentally reported. Hole carrier concentration was optimized for each condition.}
\begin{center}
\begin{tabular}{c c c c c}
\hline \hline
lattice & cubic& \multicolumn{3}{c}{orthorhombic ($Imma$)} \\
axis & $a=b=c$ & $a$ & $b$ & $c$ \\
\hline
Ca$_3$GeO & 0.05 & 0.07 & 0.06 & 0.06\\
Ba$_3$SnO & 0.06 & 0.05 & 0.05 & 0.05\\
Ba$_3$PbO & 0.08 & 0.06 & 0.06 & 0.07\\
\hline \hline
\end{tabular}
\end{center}
\end{table}

\begin{table}
\caption{\label{table:ZT_ortho2} Calculated $ZT$ values of the cubic ($Pm\bar{3}m$) and orthorhombic ($Pnma$) structures at 300 K. Calculation results are only shown for materials for which the $Pnma$ structure was experimentally reported or predicted to be stable by our calculation as presented in Fig.~\ref{fig:stab}. While the hexagonal structure was the most stable in our calculation for Ba$_3$AsN, $ZT$ values in the $Pnma$ phase are also shown for this compound because the total energies of the orthorhombic and hexagonal structures were comparable in Fig.~\ref{fig:stab}. Hole carrier concentration was optimized for each condition.}
\begin{center}
\begin{tabular}{c c c c c}
\hline \hline
lattice & cubic& \multicolumn{3}{c}{orthorhombic ($Pnma$)} \\
axis & $a=b=c$ & $a$ & $b$ & $c$ \\
\hline
Sr$_3$GeO & 0.05 & 0.07 & 0.07 & 0.08 \\
Ba$_3$GeO & 0.06 & 0.18 & 0.20 & 0.26 \\
Ca$_3$AsN & 0.07 & 0.07 & 0.07 & 0.06 \\
Sr$_3$AsN & 0.12 & 0.18 & 0.20 & 0.16 \\
Ba$_3$AsN & 0.18 & 0.15 & 0.20 & 0.16 \\
\hline \hline
\end{tabular}
\end{center}
\end{table}

\begin{figure}
\begin{center}
\includegraphics[width=8.5 cm]{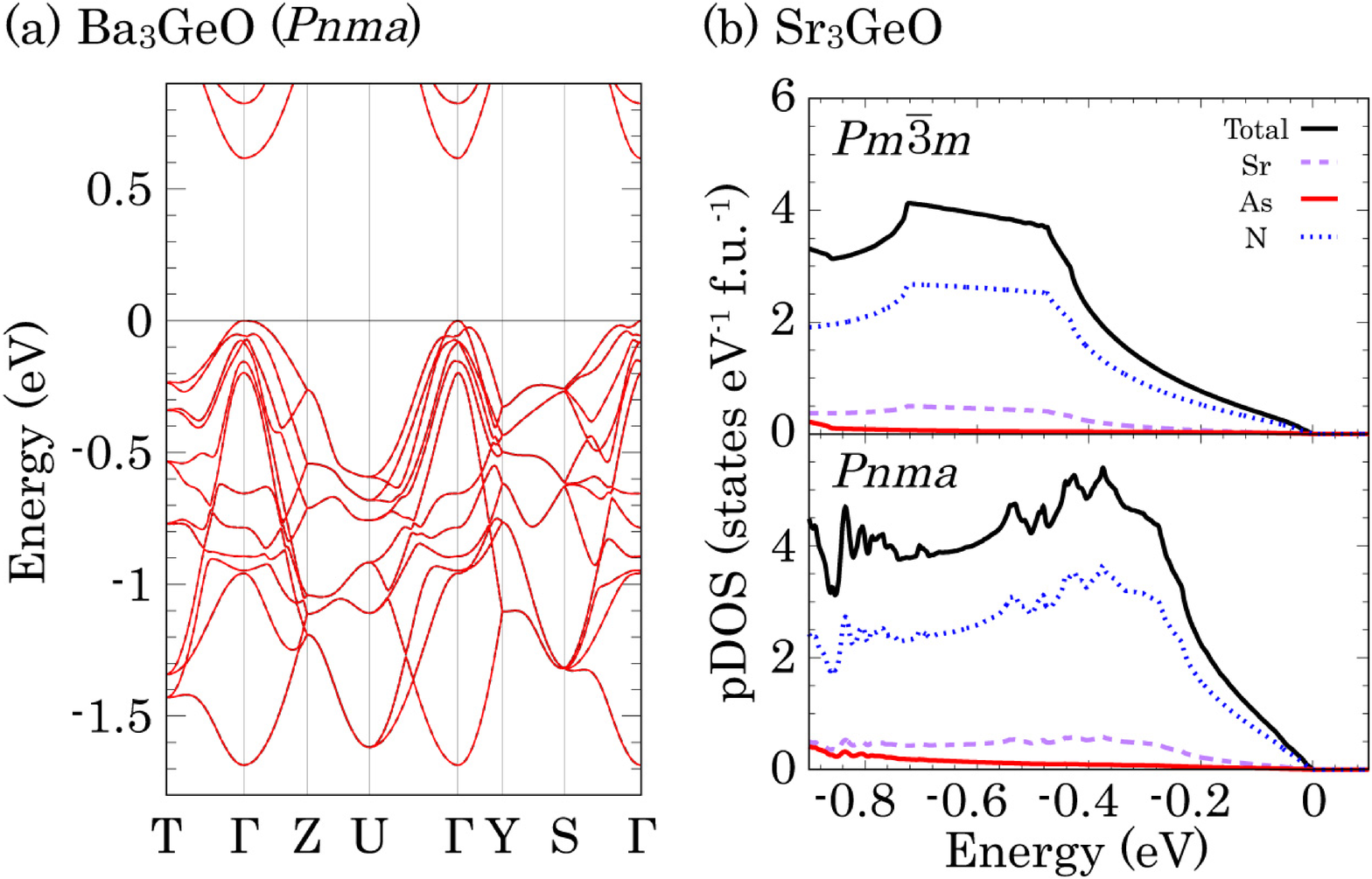}
\caption{(a) Band structure of Ba$_3$GeO in the $Pnma$ phase. Black broken and red solid lines in the band structure represent the band structures obtained with the first-principles calculation and the tight-binding model for the Wannier functions, respectively.
The $\bm{k}$-points represented with the crystal coordinate are as follows: T $= ({\bm b^*}+{\bm c^*})/2$, Z $={\bm c^*}/2$, U $=({\bm a^*}+{\bm c^*})/2$, Y $ = {\bm b^*}/2$, and S $= ({\bm a^*}+{\bm b^*})/2$. (b) DOS near the valence-band top for Sr$_3$GeO in the $Pm\bar{3}m$ (top) and $Pnma$ (bottom) phases.}
\label{fig:ortho}
\end{center}
\end{figure}

Tables~\ref{table:ZT_ortho1} and \ref{table:ZT_ortho2} present the calculated $ZT$ values for the orthorhombic ($Imma$ and $Pnma$, respectively) structures at 300 K, compared with those for the cubic ($Pm\bar{3}m$) structure.
Overall, the calculated $ZT$ values are similar between the cubic and orthorhombic phases for each compound.
However, we can see a notable increase in $ZT$ by introducing the orthorhombic distortion for two materials:
$ZT$ increases from 0.06 in $Pm\bar{3}m$ to 0.18 ($a$ axis), 0.20 ($b$ axis), and 0.26 ($c$ axis) in $Pnma$ for Ba$_3$GeO,
and from 0.12 in $Pm\bar{3}m$ to 0.18 ($a$ axis), 0.20 ($b$ axis), and 0.16 ($c$ axis) in $Pnma$ for Sr$_3$AsN.
We then focused on these two materials in the following analysis.
While Ba$_3$AsN also exhibits large $ZT$ values in the $Pnma$ phase, these values are similar to that in the cubic phase, and in addition, the total energy of the $Pnma$ phase is higher than that in the hexagonal phase as shown in Fig.~\ref{fig:stab}.
If synthesized in the $Pnma$ phase, Ba$_3$AsN can also be a promising thermoelectric material.

In Ba$_3$GeO with the $Pnma$ phase, the structure of which is shown in Fig.~\ref{fig:crys}(c), optimized lattice constants are $a=7.609$, $b=10.697$, and $c=7.481$ \AA.
The strength of the lattice distortion from the cubic lattice can be evaluated from the ratios $b(\sqrt{2}c)^{-1}=101.1 \%$ and $ac^{-1} = 101.7 \%$, both of which are $100 \%$ for the structure without the orthorhombic distortion because the $Pnma$ phase has a $\sqrt{2}\times\sqrt{2}\times2$-times enlarged unit cell compared with that for the $Pm\bar{3}m$ phase.
These structural parameters are roughly consistent with the experimental values measured at 296 K~\cite{oxides0}, $a=7.591(1)$, $b=10.728(1)$, $c=7.551(1)$ \AA, $b(\sqrt{2}c)^{-1}=100.5 \%$ and $ac^{-1} = 100.5 \%$.
The average bonding angle of O-Ba-O is 155.0$^{\circ}$ in our calculation and 158.4$^{\circ}$ in the experiment, which also show good agreement.

Figure~\ref{fig:ortho}(a) presents the band structure of Ba$_3$GeO in the $Pnma$ phase.
The most important change from that in the $Pm\bar{3}m$ phase shown in Fig.~\ref{fig:band_o}(g) is the gap opening, which is an important reason for the strong enhancement in $ZT$.
In addition, we can see an approximate but high valley degeneracy near the valence-band top in Fig.~\ref{fig:ortho}(a), which should be another cause of the high performance through the resulting large DOS.
A possibly related feature for the high valley degeneracy can be found in the $Pm\bar{3}m$ phase: the energy level of the valley at the R point is relatively close to that at the $\Gamma$ point as shown in Fig.~\ref{fig:band_o}(g). We note that the R point in the $Pm\bar{3}m$ phase is folded into the $\Gamma$ point in the $Pnma$ phase. We can also notice that the quasi-one-dimensionality of the valence band structure in the $Pm\bar{3}m$ phase, which seems comparable to that for Ca$_3$GeO, likely plays an important role for the high $ZT$ of Ba$_3$GeO in the $Pnma$ phase.

Figure.~\ref{fig:ortho}(b) presents the calculated DOS of Sr$_3$AsN in the $Pm\bar{3}m$ and $Pnma$ phases.
While both phases have a gapped band structure, DOS near the valence-band top is again enhanced by the structural distortion, which is likely to be an origin of the high $ZT$.
Because the overall shape of DOS is similar between these two phases, the increase in $ZT$ for Sr$_3$AsN is moderate compared with Ba$_3$GeO.

\begin{figure}
\begin{center}
\includegraphics[width=8.5 cm]{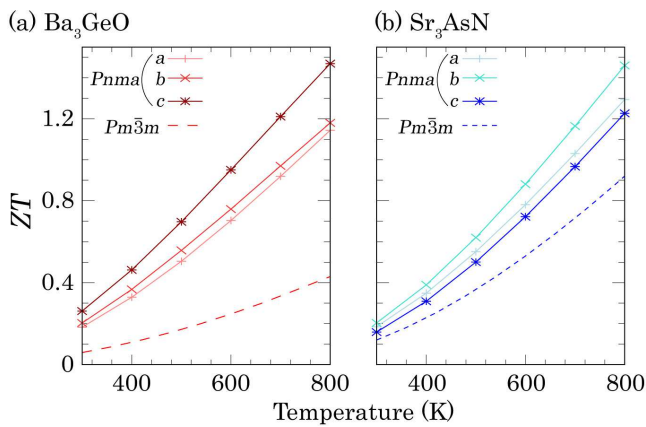}
\caption{Temperature dependence of $ZT$ for (a) Ba$_3$GeO and (b) Sr$_3$AsN. For the $Pnma$ phase, $ZT$ values for each axis are shown. The hole carrier concentration was optimized for each point.}
\label{fig:Pnma_ZT}
\end{center}
\end{figure}

We also calculated the temperature dependence of $ZT$ for Ba$_3$GeO and Sr$_3$AsN in the $Pm\bar{3}m$ and $Pnma$ phases, as shown in Fig.~\ref{fig:Pnma_ZT}.
The resulting $ZT$ in the $Pnma$ phase for each compound is comparable to the highest value in the $Pm\bar{3}m$ phase shown in Fig.~\ref{fig:ZT_n}, which was, however, obtained for Ba$_3Pn$N where in reality the hexagonal phase is likely realized.
In this sense, Ba$_3$GeO and Sr$_3$AsN with the $Pnma$ phase are the most promising candidates investigated in this study.
While there is a direction dependence of $ZT$ to some extent for both compounds in the $Pnma$ phase as shown in Fig.~\ref{fig:Pnma_ZT},
the high $ZT$ along all the directions is advantageous for technological applications.

\section{Conclusion\label{sec:sum}}

We have investigated the thermoelectric performance of the hole-doped antiperovskite oxides and nitrides by means of first-principles band structure calculation and the subsequent transport calculation based on the Boltzmann transport theory.
For the cubic $Pm\bar{3}m$ phase, we have found that Ba$_3$PbO at low temperatures (around room temperature), Ca$_3$GeO at high temperatures, and Sr$_3$SbN in the wide temperature region are promising candidates for high thermoelectric performance.
In Ba$_3$PbO, multiple Dirac cones with six-fold degeneracy and the existence of the other valleys near the $\Gamma$ and R points with relatively close energy levels enhance $ZT$.
In Ca$_3$GeO and Sr$_3$SbN, quasi-one-dimensional band dispersion originating from the orbital anisotropy of the $p$ orbitals is a key for their high performance.
When considering the other crystal structures, the hexagonal structure is not favorable at least from the perspective of the shape of the band dispersion.
However, the orthorhombic distortion toward the $Pnma$ phase sizably enhances the thermoelectric performance of Ba$_3$GeO and Sr$_3$AsN,
which are the most promising candidates within the materials investigated in this study.
For both compounds, a crucial consequence of the structural distortion is the high valley degeneracy, which is considered to enhance their thermoelectric performance.
For Ba$_3$GeO, another important role of the structural distortion is the gap opening.
Another promising candidate is Ba$_3$AsN if synthesized as the $Pnma$ phase, which is slightly unstable compared with the hexagonal phase in our calculation.
Our study will help and stimulate the experimental exploration of the high thermoelectric performance in antiperovskites oxides and nitrides, which offer a unique and fertile playground where various kinds of characteristic electronic structure take place.

\acknowledgments
Some calculations were performed using large-scale computer systems in the supercomputer center of the Institute for Solid State Physics, the University of Tokyo, and those of the Cybermedia Center, Osaka University.
This study was supported by JSPS KAKENHI (Grant No.~JP17H05481 and No.~JP19H04697) and JST CREST (Grant No.~JPMJCR16Q6), Japan.\\

\section*{Appendix: List of the optimal hole carrier density at $T=300$ K}

The optimal hole carrier density for $ZT$ at $T=300$ K is shown in Tables~\ref{tab:nopt_1}--\ref{tab:nopt_5}.

\begin{table}
\caption{\label{tab:nopt_1} Hole carrier density that maximizes the calculated $ZT$ value, $n_{\mathrm{opt}}$, at $T=300$ K for oxides $Ae_3Tt$O with the $Pm\bar{3}m$ phase.}
\begin{tabular}{c|c|c}
\hline\hline
$Ae$ & $Tt$ & $n_{\mathrm{opt}}$ ($10^{20}$ $\mathrm{cm}^{-3}$) \\
\hline
\multirow{3}{*}{Ca} & Ge & 0.5\\
 & Sn & 0.2\\
 & Pb & 0.1\\
 \cline{2-3}
\hline
\multirow{3}{*}{Sr} & Ge & 0.4\\
 & Sn & 0.2\\
 & Pb & 0.1\\
 \cline{2-3}
\hline
\multirow{3}{*}{Ba} & Ge & 6.8\\
 & Sn & 0.4\\
 & Pb & 0.5\\
 \cline{2-3}
\hline \hline
\end{tabular}
\end{table}

\begin{table}
\caption{\label{tab:nopt_2} Hole carrier density that maximizes the calculated $ZT$ value, $n_{\mathrm{opt}}$, at $T=300$ K for nitrides $Ae_3Pn$N with the $Pm\bar{3}m$ phase.}
\begin{tabular}{c|c|c}
\hline\hline
$Ae$ & $Pn$ & $n_{\mathrm{opt}}$ ($10^{20}$ $\mathrm{cm}^{-3}$) \\
\hline
\multirow{3}{*}{Ca} & As & 0.6\\
 & Sb & 0.2\\
 & Bi & 0.1\\
 \cline{2-3}
\hline
\multirow{3}{*}{Sr} & As & 0.9\\
 & Sb & 1.2\\
 & Bi & 0.2\\
 \cline{2-3}
\hline
\multirow{3}{*}{Ba} & As & 4.0\\
 & Sb & 3.8\\
 & Bi & 4.3\\
 \cline{2-3}
\hline \hline
\end{tabular}
\end{table}

\begin{table}
\caption{\label{tab:nopt_3} Hole carrier density that maximizes the calculated $ZT$ value, $n_{\mathrm{opt}}$, at $T=300$ K for materials with the $P6_3/mmc$ phase.
In this table, $n_{\mathrm{opt}}$ for each axis is presented.}
\begin{tabular}{ccc}
\hline\hline
 & \multicolumn{2}{c}{$n_{\mathrm{opt}}$ ($10^{20}$ cm$^{-3}$)} \\
& $a=b$ & $c$ \\
\hline
Ba$_3$AsN & 13 & 28 \\
Ba$_3$SbN & 0.7 & 22 \\
Ba$_3$BiN & 0.3 & 27 \\
\hline \hline
\end{tabular}
\end{table}

\begin{table}
\caption{\label{tab:nopt_4} Hole carrier density that maximizes the calculated $ZT$ value, $n_{\mathrm{opt}}$, at $T=300$ K for materials with the $Imma$ phase.
In this table, $n_{\mathrm{opt}}$ for each axis is presented.}
\begin{tabular}{cccc}
\hline\hline
 & \multicolumn{2}{c}{$n_{\mathrm{opt}}$ ($10^{20}$ cm$^{-3}$)} \\
& $a$ & $b$ & $c$ \\
\hline
Ca$_3$GeO & 0.2  & 0.8 & 0.7 \\
Ba$_3$SnO & 0.2 & 0.2 & 0.2 \\
Ba$_3$PbO & 0.4 & 0.3 & 0.4 \\
\hline \hline
\end{tabular}
\end{table}

\begin{table}
\caption{\label{tab:nopt_5} Hole carrier density that maximizes the calculated $ZT$ value, $n_{\mathrm{opt}}$, at $T=300$ K for materials with the $Pnma$ phase.
In this table, $n_{\mathrm{opt}}$ for each axis is presented.}
\begin{tabular}{cccc}
\hline\hline
 & \multicolumn{2}{c}{$n_{\mathrm{opt}}$ ($10^{20}$ cm$^{-3}$)} \\
& $a$ & $b$ & $c$ \\
\hline
Sr$_3$GeO & 1.3 & 0.6 & 0.7 \\
Ba$_3$GeO & 1.7 & 1.6 & 1.5 \\
Ca$_3$AsN & 0.6 & 1.5 & 1.2 \\
Sr$_3$AsN & 2.5 & 2.2 & 3.0 \\
Ba$_3$AsN & 9.1 & 3.9 & 9.6 \\
\hline \hline
\end{tabular}
\end{table}


\begin{thebibliography}{999}
\bibitem{note_phonon} As a related study, a recent first-principles calculation~\cite{Haque5} using the GGA-PBE functional~\cite{PBE} without SOC reported that Sr$_3$AsN with the $Pm\bar{3}m$ phase has no imaginary phonon frequency.
\end{thebibliography}
\end{document}